# *In vitro* toxicity and uptake of magnetic nanorods

**M. Safi, S. Clowez, A. Galimard and J.-F. Berret**
Matière et Systèmes Complexes, UMR 7057 CNRS Université Denis Diderot Paris-VII, Bâtiment Condorcet, 10 rue Alice Domon et Léonie Duquet, 75205 Paris (France)

E-mail: jean-francois.berret@univ-paris-diderot.fr

**Abstract.** In this paper we investigate the internalization and cytotoxicity of nanostructured materials having the form of elongated rods, with diameter of 200 nm and lengths 1 – 10 μm. The rods were made from the controlled aggregation of sub-10 nm iron oxide nanoparticles. Recently, we have shown that the nanorods inherited the superparamagnetic property of the particles. These rods can actually be moved by the application of an external magnetic field. Here we evaluate the *in vitro* toxicity of the magnetic nanorods by using MTT assays on NIH/3T3 mouse fibroblasts. The toxicity assays revealed that the nanorods are biocompatible, as exposed cells remained 100% viable relative to controls over a period of a few days. Optical microscopy allow to visualize the rods inside the cells and to determine their number per cell. Roughly 1/3 of the total incubated rods were uptaken by the fibroblasts.

Inorganic nanomaterials and particles with enhanced optical, mechanical or magnetic attributes are currently being developed for a wide scope of applications, including catalysis, photovoltaics, coating and nanomedicine [1]. More generally, nanomaterials of different shapes and sizes are regarded as promising tools at the scale of the cell for manipulation, diagnostic and therapy. Interactions between nanomaterials and living organisms were also investigated extensively because of the concerns raised by their potential toxicity. As shown in several reviews [2, 3], the risk assessment of nanomaterials towards living cells and tissues have not been fully evaluated. Among the wide variety of nanomaterials available, magnetic nanowires have received considerable attention [4-8]. Nanowires are anisotropic colloidal objects with submicronic diameters and lengths in the range 1 – 100 μm. In specific applications such as cell separation, ferromagnetic nickel nanowires made by electrodepostion were shown to outperform magnetic beads of comparable volume [5]. However, one major drawback encountered with nickel or iron nanowires is that these objects carry a permanent magnetic moment, and are thus susceptible to aggregate in solution because of magnetic dipolar attraction [4].

In the present paper, we exploit a simple and versatile waterborne synthesis process to generate magnetic nanorods [9, 10]. Highly persistent rods of diameters around 200 nm and lengths comprised between 1 μm and 30 μm were fabricated by controlling the assembly of sub-10 nm iron oxide nanoparticles. These magnetic nanorods are different from the ferrimagnetic electrodeposited wires mentioned previously. The rods are superparamagnetic *i.e.* they do not carry a permanent magnetic moment, which prevents their spontaneous aggregation in a dispersion [9, 11]. Since these rods are designed to be used as micromechanical tools at the cellular level, their interactions with living cells need to be investigated. Here we explore their interactions with NIH/3T3 mouse fibroblasts and show that the rods were internalized in the cellular matrix. Toxicity assays show that the rods do not display acute toxicity towards the cells on a short-term basis. In this study, the effects of the rods on the fibroblasts and those of the nanoparticles from which they are made were also compared.

## 1. Materials and Methods

### 1.1. Chemicals, synthesis and characterization

The synthesis of superparamagnetic nanoparticles (NP) used the technique of 'soft chemistry' based on the polycondensation of metallic salts in alkaline aqueous media elaborated by R. Massart [12]. This technique resulted in the formation of magnetite ($Fe_3O_4$) NPs of sizes comprised between 4 and 15 nm. Magnetite nanocrystals were further oxidized into maghemite ($\gamma$-$Fe_2O_3$) and sorted according to their size. Adding large amount of nitric acid to the dispersions down to pH 0.5 initiated a liquid-gas type phase separation, resulting in coexisting dilute and concentrated phases of $\gamma$-$Fe_2O_3$ particles. The concentrated phase which contained the largest particles was separated by magnetic sedimentation. The process was repeated 5 times, reducing the initial polydispersity from 0.4 to 0.2. In the synthesis conditions, the magnetic dispersions were stabilized by electrostatic interactions arising from the native cationic charges at the surface of the particles. Vibrating sample magnetometry was performed to characterize the size distribution of the maghemite cores [13, 14]. Fig. 1 shows the evolution of the macroscopic magnetization M(H) normalized by its saturation value $M_S$ for the present $\gamma$-$Fe_2O_3$ batch. Here, $M_S = \phi m_S$, where $m_S$ is the volumetric magnetization of maghemite ($m_S = 2.9 \times 10^5$ A m$^{-1}$).

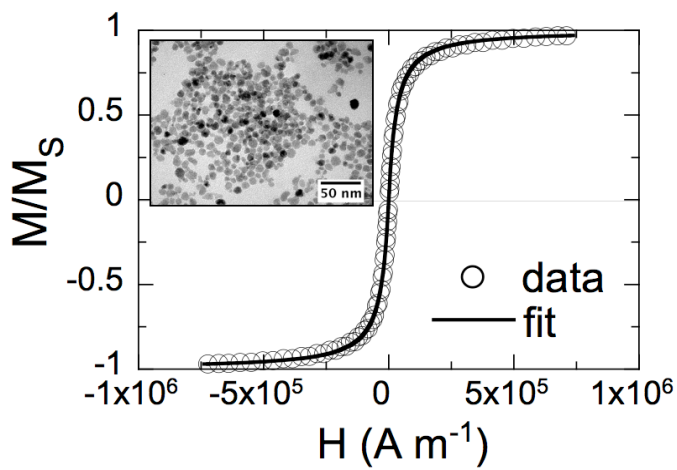

**Figure 1**: Magnetic field dependence of the macroscopic magnetization M(H) of a $\gamma$-$Fe_2O_3$ dispersion normalized by its saturation value $M_S$. The solid curve was obtained using the Langevin function convoluted with a log-normal distribution with median diameter $D_0 = 6.7$ nm and polydispersity $s_{NP} = 0.21$. Inset : Transmission electron microscopy of iron oxide at the magnification of ×120000.

The solid curve in Fig. 1 was obtained using the Langevin function convoluted with a log-normal distribution function of the particle size. The parameters of the distribution are the median diameter $D_0 = 6.7$ nm and the polydispersity $s_{NP} = 0.21$. The polydispersity was defined as the ratio between the standard deviation and average diameter. A transmission electron microscopy image obtained from the same $\gamma$-$Fe_2O_3$ dispersion is illustrated in the inset of Fig.1. The iron oxide NPs exhibited compact and spherical structures, with size and polydispersity in excellent agreement with the VSM data [14].

In order to improve their colloidal stability, the cationic particles were coated by poly(acrylic acid) oligomers. Poly(sodium acrylate), the salt form of poly(acrylic acid) with a molecular weight $M_W =$ 2000 g mol$^{-1}$ and a polydispersity of 1.7 was purchased from Sigma Aldrich and used without further purification. In order to adsorb polyelectrolytes on the surface of the nanoparticles, we followed the *Precipitation–Redispersion* protocol, as described elsewhere [14, 15]. The precipitation of the iron oxide dispersions by PAA$_{2K}$ was performed in acidic conditions (pH 2). The precipitate was separated from the solution by centrifugation and its pH was increased by the addition of ammonium hydroxide. The precipitate redispersed spontaneously at pH 7–8, yielding a clear solution that contained the polymer coated particles. The hydrodynamic sizes found in coated NPs were 6 nm above that of the bare particles, indicating the presence of a 3 nm PAA$_{2K}$ brush surrounding the particles. This simple technique allowed us to produce large quantities of coated particles (> 1 g of oxides) within a relatively short time (< 1 h) [16]. The fabrication of the rods was performed using these coated particles, noted PAA$_{2K}$–γ-Fe$_2$O$_3$ in the following.

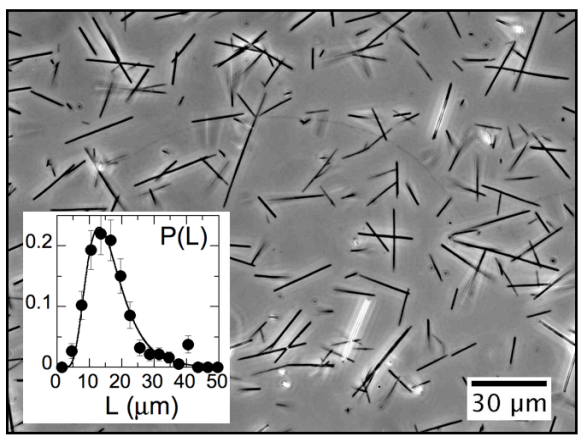

**Figure 2** : Phase-contrast optical microscopy images (40×) of a dispersion of nanostructured rods made from 6.7 nm γ-Fe$_2$O$_3$ particles in absence of magnetic field. Inset : Length distribution of the rods. The distribution was found to be log-normal with median length and polydispersity noted L$_0$ = 15 μm and s$_R$ = 0.40 respectively.

The fabrication of the nanorods is based on the NPs co-assembly in the presence of a magnetic field. The adopted strategy involved in a first step the preparation of two separate 1 M NH$_4$Cl solutions containing respectively *i)* the anionic iron oxide NPs (PAA$_{2K}$–γ-Fe$_2$O$_3$) and *ii)* the PTEA$_{11K}$-*b*-PAM$_{30K}$ diblock copolymers. These copolymers (PTEA-*b*-PAM stands for poly(trimethylammonium ethylacrylate)-*b*-poly(acrylamide)) were synthesized by MADIX® controlled radical polymerization which is a Rhodia patented process [17]. The initial concentrations were adjusted to c = 0.1 wt. % for both NPs and copolymers. In a second step, the two solutions were mixed with each other and it was checked by dynamic light scattering that the two components remained dispersed [9, 11]. In a third step, the ionic strength of the mixture was progressively diminished by dialysis in the presence of an external magnetic field (0.1 T). Dialysis was performed against de-ionized water using a Slide-a-Lyzer® cassette with MWCO of 10 kD (Thermo Scientific). The volume of the dialysis bath was 300 larger than that of the samples. The electrical conductivity (Cyberscan PC6000) of the dialysis bath was measured during the ion exchange and served to monitor the desalting kinetics. In the conditions described here, the whole process reached a stationary and final state within one hour. Once the ionic strength of the bath reached its stationary value, (3×10$^{-3}$ M), the external magnetic field was removed and the dispersions were studied by optical microscopy (Fig. 2). In order to determine the length distribution of the rods, pictures were digitized and treated by the ImageJ software (http://rsbweb.nih.gov/ij/). The inset of Fig. 2 displays the length distribution function that was determined from a panel of 250 objects. It was found to be log-normal with a median length of L$_0$ = 15 µm and a polydispersity $s_R = 0.4$ (continuous line). Rods were further sonicated to reduce their average length down to 3 µm.

*1.2. Experimental Methods*

*1.2.1. Cell culture*
NIH/3T3 fibroblast cells from mice were grown as a monolayer in Dulbecco's modified Eagle's medium (DMEM) with high glucose (4.5 g L$^{-1}$) and stable glutamine (PAA Laboratories GmbH, Austria). This medium was supplemented with 10% fetal bovine serum (FBS) and 1% penicillin/streptomycin (PAA Laboratories GmbH, Austria), referred to as cell culture medium. Exponentially growing cultures were maintained in a humidified atmosphere of 5% $CO_2$ and 95% air at 37°C, and under these conditions the plating efficiency was 70 – 90% and the cell duplication time was 12–14 h. Cell cultures were passaged twice weekly using trypsin–EDTA (PAA Laboratories GmbH, Austria) to detach the cells from their culture flasks and wells. The cells were pelleted by centrifugation at 1200 rpm for 5 min.

*1.2.2. Transmission optical Microscopy*
For optical microscopy observations, phase-contrast images of the cells containing rods were acquired on an IX71 inverted microscope (Olympus) equipped with 40× and 60× objectives. $2\times10^4$ NIH/3T3 fibroblasts cells were first seeded onto a 96-well plate for 24 h prior being incubated with nanorods. µl-aliquots containing nanorods were added to the supernatant. The incubation of the rods lasted 24 hours. The third day, the excess medium was removed and the cells were washed with PBS solution (with calcium and magnesium, Dulbecco's, PAA Laboratories), trypsinized and centrifuged. The cell pellets were resuspended in DMEM. For optical microscopy, 20 µl of the previous cell suspension were deposited on a glass plate and sealed into to a Gene Frame® (Abgene/Advanced Biotech) dual adhesive system. The sample was then left for 4 h in the incubator to let the cells adhere onto the glass plate. Images were observed using a Photometrics Cascade camera (Roper Scientific) and Metaview software (Universal Imaging Inc.) as acquisition system.

*1.2.3. MILC protocol*
UV-visible spectrometry was performed in the MILC protocol (**M**ass of metal **I**nternalized/Adsorbed by **L**iving **C**ells) which consist in the measurement of the mass of nanoparticles incorporated into living cells. The quantity to be determined is the mass of iron expressed in the unit of picogram per cell. Cells were seeded onto 3.6 cm Petri dishes, and incubated until reaching 60% confluence and then incubated with nanomaterials, including the nanorods at different concentrations for 24 h. After the incubation period, the supernatant was removed and the layer of cells washed thoroughly with PBS. The cells were then trypsinised and mixed again with white DMEM without serum. 20 $\mu$l-aliquots were taken up for counting using a Malassez plate. The cells were finally centrifuged and pellets were dissolved in 35% HCl. When incubated with γ-$Fe_2O_3$, the HCl dissolved NIH/3T3 exhibited a yellow color. The absorbance of the liquid was studied using a Variant spectrophotometer (Cary 50 Scan), and it was compared to a calibrated reference. The reference was obtained by dissolving directly γ-$Fe_2O_3$ dispersions of known concentrations into 35% HCl.

*1.2.4. MTT toxicity assays*

MTT assays were performed with $PAA_{2K}$-coated iron oxide nanoparticles and nanorods for metal molar concentrations [Fe] between 10 μM and 10 mM. To avoid contamination from synthesis residues and free ions, nanoparticle and nanorod dispersions were dialyzed against DI-water which pH was adjusted to 8 by addition of sodium hydroxide (Spectra Por 2 dialysis membrane with MWCO 12 kD). Cells were seeded into 96-well microplates, and the plates were placed in an incubator overnight to allow for attachment and recovery. Cell densities were adjusted to $2\times10^4$ cells per well (200 μl). After 24 h, the nanoparticles and nanorods were applied directly to each well using a multichannel pipette. Cultures were incubated in triplicate for 24 h at 37°C. The MTT assay depends on the cellular reduction of MTT (3- (4,5-dimethylthiazol-2-yl)-2,5-diphenyl tetrazolium bromide, Sigma Aldrich Chemical) by the mitochondrial dehydrogenase of viable cells forming a blue formazan product which can be measured spectrophotometrically [18]. MTT was prepared at 5 mg ml$^{-1}$ in PBS (with calcium and magnesium, Dulbecco's, PAA Laboratories) and then diluted 1–5 in medium without serum and without Phenol Red. After 24 h of incubation with nanoparticles, the medium was removed and 200 μl of the MTT solution was added to the microculture wells. After 4 h incubation at 37°C, the MTT solution was removed and 100 μl of 100% DMSO was added to each well to solubilize the MTT–formazan product. The absorbance at 562 nm was then measured with a microplate reader (Perkin Elmer). Prior to the microplate UV–vis spectrometry, MTT assays without particles were carried out with cell populations ranging from $5\times10^3$ to $5\times10^5$ cells.

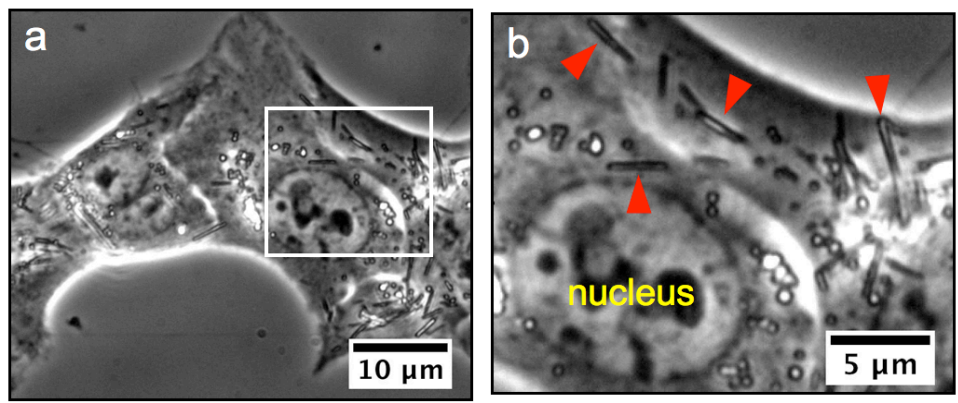

**Figure 3.** a) Phase contrast microscopy images of NIH/3T3 fibroblasts cells treated with nanorods for 24 h at a concentration of 30 rods/cell. b) Zoom of the previous image to emphasize the perinuclear region of a unique cell. Rods of various lengths and orientations are indicated by arrows.

## 2. Results and discussion

*2.1. Nanorods are internalized by the cells*
Fig. 3a shows a cluster of cells that were incubated with magnetic nanorods of average length 10 μm and at 30 rods per cell. In this work, the nanorod concentration was defined by the ratio of the number of rods incubated per cell. To allow comparison with data on $\gamma$-$Fe_2O_3$ NPs or with data from the literature, it was also expressed in terms of iron molar concentration [Fe] (30 rods per cell corresponds to [Fe] = 0.15 mM). The cells were found to maintain their morphology and adherence properties after a 24 h exposure. Fig. 3b displays a zoom of the previous image and emphasizes the perinuclear region of a single cell. There, elongated threads with lengths comprised between 1 to 7 μm and with random orientations are observed (arrows). The location of the rods into the cytoplasm and not outside in the supernatant or attached at the cellular membrane was confirmed by following the their Brownian motions. The orientational fluctuations of the rods were recorded as a function of the time and compared with those of rods dispersed in water. It was found that the rotational diffusion constant was much slower for the internalized objects, in good agreement with recent estimations of the viscosity of the intracellular matrix [19]. These results also suggest that the internalization of nanorods was optimized when their length corresponded to the average size of the adherent cells [5].

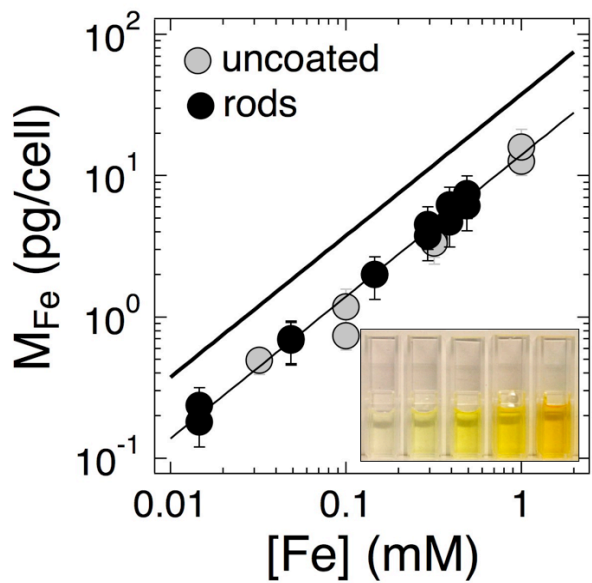

**Figure 4** : Amount of internalized or adsorbed iron oxide $M_{Fe}$ (pg/cell) determined following the MILC protocol (see text for details). The upper thick straight line depicts the maximum amount of iron that can be uptaken by the cells for a molar concentration [Fe] in the supernatant. Grey bullets : uncoated $\gamma\text{-}Fe_2O_3$; closed bullets : nanorods ($L_0 = 3$ μm, $s_R = 0.5$). Inset : UV-cells containing increasing concentrations of iron oxide dissolved in HCl

*2.2. Quantitative determination of internalized iron oxide*
The amount of internalized or adsorbed iron oxide was determined following the protocol MILC (**M**ass of metal **I**nternalized/Adsorbed by **L**iving **C**ells) which makes use of UV spectrophotometry to calculate the iron concentration from pelleted cells dissolved in concentrated HCl. The results ($M_{Fe}$ *versus* [Fe]) were expressed in picograms of iron per cell. Here, [Fe] denotes the iron oxide molar concentration in the supernatant during the 24 h of incubation. Fig. 4 compares the [Fe]-dependences of internalized/adsorbed amounts for uncoated NPs and for the nanorods ($L_0 = 3$ μm, $s_R = 0.5$). In the inset, UV-cells containing increasing concentrations of iron oxide dissolved in HCl are shown for illustration. The thick straight line in Fig. 4 depicts the maximum amount of iron that could be uptaken by the cells. For instance, $3\times10^6$ of fibroblasts exposed to a [Fe] = 1 mM solution, the maximum amount lies at 37 pg/cell. For the uncoated particles and for the nanorods, $M_{Fe}$ increases linearly and represents about 35% of the maximum value. Optical microscopy carried out 24 h after incubation and thorough washing shown that none of the rods were adsorbed at the cell membranes, leading to the conclusion that the $M_{Fe}$-data in Fig. 4 represent the amount of internalized iron under the form of nanorod. For uncoated $\gamma\text{-}Fe_2O_3$, the large quantities detected up to [Fe] = 10 mM were explained by the precipitation of the particles in the culture medium. This precipitation produced large and compact clusters in the micrometer range (1 – 20 μm) [16] that for the most part were adsorbed on the plasma membrane. Thorough washing with PBS buffer did not desorb these aggregates, probably because of their positive surface charges.

*2.3. Toxicity assays*

Cytoxicity studies determine the cellular alterations or damages of vital functions induced by xenobiotics. MTT viability assays were conducted on NIH/3T3 cells for the uncoated $\gamma$-$Fe_2O_3$ particles ([Fe] = 10 $\mu$M - 10 mM) and for the nanorods ($L_0$ = 3 $\mu$m, [Fe] = 0.1 $\mu$M - 1 mM). The comparison between this two types of nanomaterials aimed to identify the role of size and morphology on the cell survival. A rod concentration of 1 mM corresponded to 100 rods per cell. Exposure times were set at 24 h. As shown in Figs. 5 for both studies, the viability remained at a 100% level within experimental accuracy. These findings indicate a normal mitochondrial activity for the cultures tested. The results on the precipitated $\gamma$-$Fe_2O_3$ are in good agreement with earlier reports [20, 21]. As mentioned before, for this system most of the NPs were precipitated into large clusters that were not internalized because of their size (> 10 $\mu$m), but rather adsorbed on the plasma membranes. For the nanorods, the data also confirm those obtained recently on parent micron-size materials, such as silica [22] or carbon nanotubes [23]. In conclusion, the MTT assays show conclusive evidences and ensure the suitability of the rods for biomedical and biophysical applications.

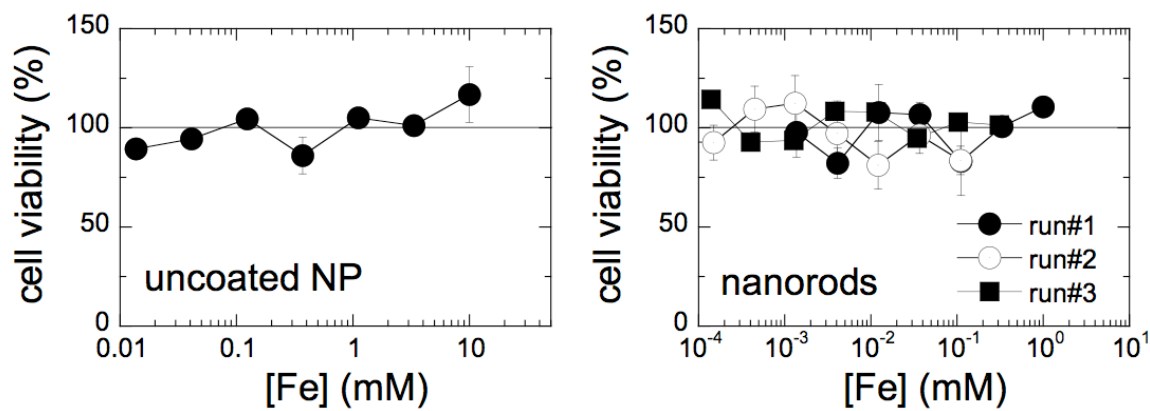

**Figure 5.** MTT (3-(4,5-dimethylthiazol-2-yl)-2,5-diphenyl tetrazolium bromide) viability assays conducted on NIH/3T3 cells incubated during 24 h with a) uncoated $\gamma$-$Fe_2O_3$ NPs and b) 3 $\mu$m-nanorods. In this work, the nanorod concentration was defined by the ratio of the number of rods incubated per cell. To allow comparison with data from the literature, it was also expressed in terms of iron molar concentration [Fe].

## 3. Conclusion

In this paper, we have evaluated the interactions and toxicity of magnetic nanorods with respect to NIH/3T3 mouse fibroblasts. Magnetic nanorods are a particularly promising class of nanomaterials since they can be used in confined geometries for cell manipulation, microrheology and microfluidics. The magnetic properties of the rods were inherited from the iron oxide particles, and allow to rotate the rods in a propeller-like motion by the application of an external field [9, 10]. Our approach with cell culture consisted to show that the rods were internalized and biocompatible with living fibroblasts. Concerning the first point, quantitative measurements of internalized amounts revealed that 3 $\mu$m-nanorods were uptaken in large proportion, typically 35% of the initial seeded quantity. Direct visualization of the rods inside the cytoplasm by optical microscopy also confirmed these conclusions : a large number of incubated rods were able to cross the cell plasma membranes. In the future, we plan to use transmission electron microscopy to identify if the rods are located in intracellular compartments or directly dispersed in the cytosol, and to have a better insight about the portals of entry into the cells. From these initial studies, it is concluded that the iron oxide based nanorods can be used safely with living cells, e.g. as microtools for *in vitro* and *in vivo* applications.

**Acknowledgments**
This research was supported in part by Rhodia (France), by the Agence Nationale de la Recherche under the contracts BLAN07-3_206866 and ANR-09-NANO-P200-36, by the European Community through the project : "NANO3T—Biofunctionalized Metal and Magnetic Nanoparticles for Targeted Tumor Therapy", project number 214137 (FP7-NMP-2007-SMALL-1) and by the Région Ile-de-France in the DIM framework related to Health, Environnement and Toxicology (SEnT).